\documentclass[lettersize,journal]{IEEEtran}
\usepackage{amsmath,amsfonts}
\usepackage{algorithmic}
\usepackage{algorithm}
\usepackage{threeparttable}
\usepackage{array}
\usepackage[caption=false,font=normalsize,labelfont=sf,textfont=sf]{subfig}
\usepackage[dvipsnames]{xcolor}
\usepackage{textcomp}
\usepackage{stfloats}
\usepackage{url}
\usepackage{verbatim}
\usepackage{graphicx}
\usepackage{cite}
\usepackage{multirow}
\usepackage{balance}
\usepackage{placeins}
\usepackage{caption}
\begin{document}

\title{GAP-LA: \underline{G}PU-\underline{A}ccelerated \underline{P}erformance-Driven  \underline{L}ayer \underline{A}ssignment}

\author{Chunyuan Zhao, Zizheng Guo, Zuodong Zhang, Yibo Lin,~\IEEEmembership{Member,~IEEE}
 \thanks{
    This project is supported in part by Natural Science Foundation of Beijing, China (Grant No. Z230002), National Science and Technology Major Project (2021ZD0114702), and the 111 project (B18001).
    
    C. Zhao is with the School of Integrated Circuits, Peking University, Beijing 100871, China.
    
    Z. Guo is with the School of Integrated Circuits, Peking University, Beijing 100871, China and Institute of Electronic Design Automation, Peking University, Wuxi 221000, China.

    Z. Zhang is with Institute of Electronic Design Automation, Peking University, Wuxi 221000, China.

    Y. Lin is  with the School of Integrated Circuits, Peking University, Beijing 100871, China, Institute of Electronic Design Automation, Peking University, Wuxi 221000, China, and Beijing Advanced Innovation Center for Integrated Circuits, Beijing 100871, China.
    Corresponding author: Yibo Lin (yibolin@pku.edu.cn)
}
}
\markboth{}
{Shell \MakeLowercase{\textit{et al.}}: A Sample Article Using IEEEtran.cls for IEEE Journals}

\maketitle

\begin{abstract}
    Layer assignment is critical for global routing of VLSI circuits. It converts 2D routing paths into 3D routing solutions by determining the proper metal layer for each routing segments to minimize congestion and via count.
    As different layers have different unit resistance and capacitance, layer assignment also has significant impacts to timing and power. 
    With growing design complexity, it becomes increasingly challenging to simultaneously optimize timing, power, and congestion efficiently. 
    Existing studies are mostly limited to a subset of objectives.
    
    In this paper, we propose a GPU-accelerated performance-driven layer assignment framework, \texttt{GAP-LA}, for holistic optimization the aforementioned objectives. 
    Experimental results demonstrate that we can achieve 0.3\%-9.9\% better worst negative slack (WNS) and 2.0\%-5.4\% better total negative slack (TNS) while maintaining power and congestion with competitive runtime compared with ISPD 2025 contest winners, especially on designs with up to 12 millions of nets.
    We also apply our method to the 2D net topology of ISPD 2025 contest winners.
    Compared with their original flow, we achieve 8.5\%-18.3\% better WNS and 2.0\%-5.1\% better TNS.
\end{abstract}




\section{Introduction}
\label{sec:intro}

\begin{figure}[tb]
    \centering
    \includegraphics[width=0.95\linewidth]{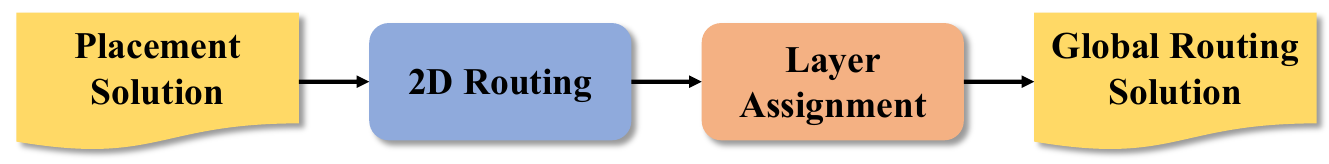}
    \caption{Typical 2D global routing flow~\cite{Xu09FastRoute4.0}.}
    \label{fig:GR}
\end{figure}

Global routing is an important step in modern very-large-scale-integrated (VLSI) physical design flow.
It partitions the layout into a set of GCells and finds a GCell-level routing solution of each signal net.
The solution serves as early routing planning for the detailed routing stage and offers fast metric feedback such as congestion for the placement stage.
Figure~\ref{fig:GR} presents a typical 2D global routing flow, which can be divided into 2 phases: 2D routing and layer assignment.
2D routing decides the connections between components while optimizing wirelength and congestion by projecting 3D routing resources into a 2D resource model.
After that, layer assignment is invoked to assign those 2D routing segments to proper layers to generate the expected 3D global routing solution while minimizing congestion and via count.

To achieve better design performance, layer assignment needs to optimize for timing and power instead of those indirect metrics.
Timing is determined by cell delay and net delay, which are highly correlated to the parasitic resistance and capacitance of nets.
Power is correlated to the parasitic capacitance of nets.
As unit resistance and capacitance of low metal layers can be several times larger than high metal layers, layer assignment can have significant impacts to timing and power. 

Prior works on layer assignment can be roughly categorized into congestion-driven and delay-driven approaches.
Congestion-driven layer assignment focuses on net ordering, escaping local optima strategies and refinement process to minimizing congestion.
Lee and Wang~\cite{Lee08LayerAssign} design net ordering strategy considering net wirelength, number of pins and average density.
They also apply a dynamic programming method and an examination method to avoid violating congestion constraints.
In~\cite{Lee09}, they take via capacity proposed by~\cite{Hsu08ViaCapacity} and bending point cost into account for a better congestion in more accurate resource model.
FastRoute~\cite{Xu09FastRoute4.0} further introduces an edge order strategy in layer assignment process.
\cite{Liu11NegoLayerAssign, Dai12NCTU-GR} use negotiation-based method to penalize congestion region to escape local optima and ~\cite{Dai12NCTU-GR} introduces layer shifting method to refine after rip-up and re-assign process.
To avoid solution space being restricted by net ordering, COALA~\cite{Jiang23COALA} assigns wires concurrently.
\cite{Liu20CUGR, Lin22Superfast, Lin23GAMER, Liu23EDGE} integrate layer assignment into 2D routing to optimize congestion in 3D resource model directly.

Delay-driven layer assignment focuses on minimizing maximal or average delay of routing segments under congestion constraints.
~\cite{Yu15TILA, Liu16Incre} prioritize nets with large delay and formulate the delay optimization problem as an ILP problem.
~\cite{Han17DelayDrivenAdv, Zhang20MiniDelay, Jiang22LASVR, Liu22TALA} take coupling effect, slew and non-default-rule routing into account to better handle delay optimization problem in advanced technology node.
They also apply edge weighting strategy to optimize maximal and average delay of sinks effectively.
It usually takes around ten minutes for those CPU layer assignment frameworks to handle the design with around ten million pins and one million GCells.

\begin{figure}[b]
    \centering
    \includegraphics[width=0.8\linewidth]{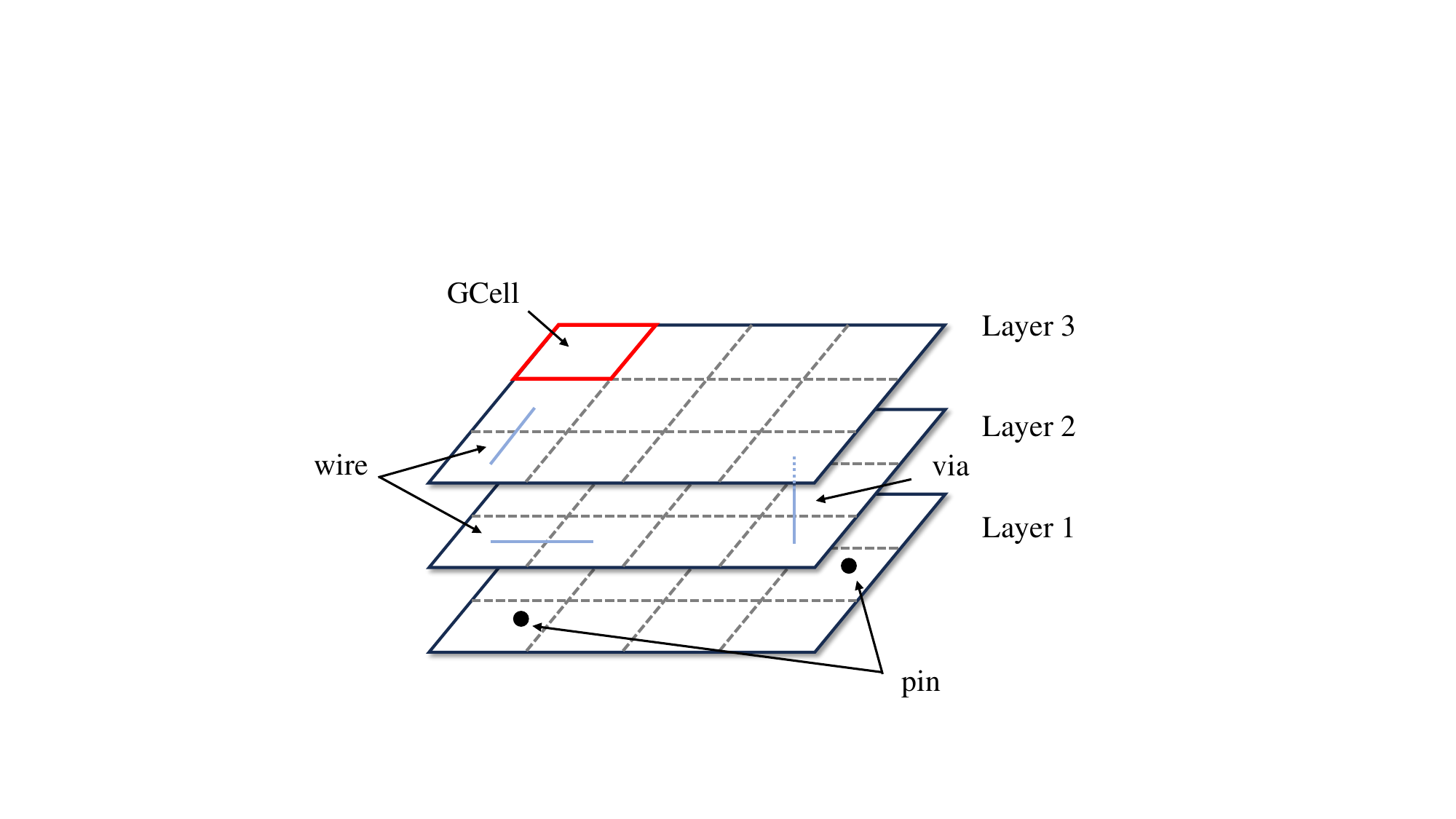}
    \caption{GCell grid graph in global routing.}
    \label{fig:GCell graph}
\end{figure}

However, the aforementioned methods still face challenges, such as insufficient runtime efficiency in modern large-scale designs and the neglect of direct optimization of timing metrics like TNS and WNS.
To optimize TNS and WNS, layer assignment is desired to have a comprehensive analysis across the entire circuit design and introduce effective optimization strategy instead of only considering maximal and average delay.
As design complexity and scale grows, layer assignment on CPU is facing runtime efficiency problem.
In recent years, GPU acceleration techniques have demonstrated significant potential in physical design (PD). 
Leveraging the massive parallel computing power of GPUs, the runtime of many traditional problems has been reduced by tens of times, particularly in the placement~\cite{Lin21DreamPlace, Lin20DreamPlace2.0, Gu20DreamPlace3.0, Liao22DreamPlace4.0, Liu22Xplace, Mai25LEGALM} and routing~\cite{Liu22FastGR, Lin23GAMER, Lin22Superfast, Lin25Instant, Zhao25helemgr} stages.
However, very few research has explored efficient optimization of timing and power in layer assignment with GPU acceleration. 

To tackle the performance optimization of such challenging cases, we propose \texttt{GAP-LA}, which utilizes different strategies and GPU-accelerated layer assignment kernels to achieve high-quality 3D global routing solution.

We summarize our main contributions as follows:
\begin{itemize}
    \item We present a performance-driven layer assignment framework \texttt{GAP-LA} with GPU acceleration. In this framework, we consider net delay and net capacitance to optimize timing and power.
    \item We propose a net ordering strategy considering timing information reported by static timing analysis to prioritize the timing critical nets for explicit WNS and TNS optimization effectively.
    \item We propose a look ahead strategy to handle delay cost in tree dynamic programming-based layer assignment method and fully accelerate the assignment process with GPU.
\end{itemize}

Experimental results show that, compared with the top-3 winners in ISPD25 Contest~\cite{ISPD2025}, our method achieve 0.3\%-9.9\% better WNS and 2.0\%-5.4\% better TNS while maintaining comparable power consumption and congestion.

The rest of the paper is organized as follows.
Section~\ref{sec:prelim} introduces the background of our performance-driven layer assignment algorithms and the problem formulation.
Section~\ref{sec:algo} explains our strategies and GPU kernels used in \texttt{GAP-LA}.
Section~\ref{sec:exp} validates our approach with comprehensive experimental results.
Section~\ref{sec:conclu} concludes our paper.
\section{Preliminaries}
\label{sec:prelim}

\begin{figure}[tb]
    \centering
    \subfloat[]{
        \includegraphics[width=0.35\linewidth]{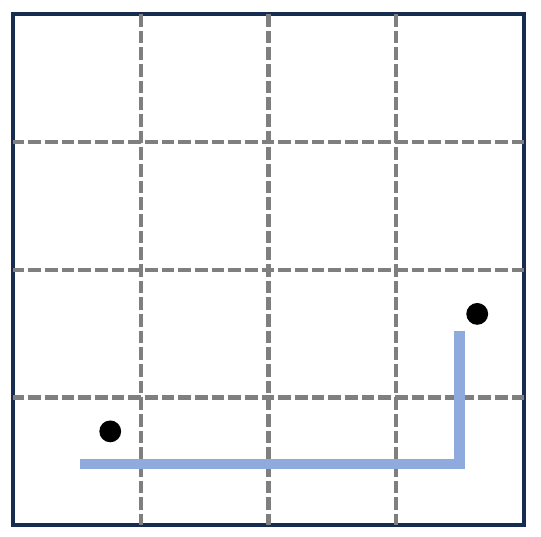}
        \label{fig:2d}
    }
    \subfloat[]{
        \includegraphics[width=0.55\linewidth]{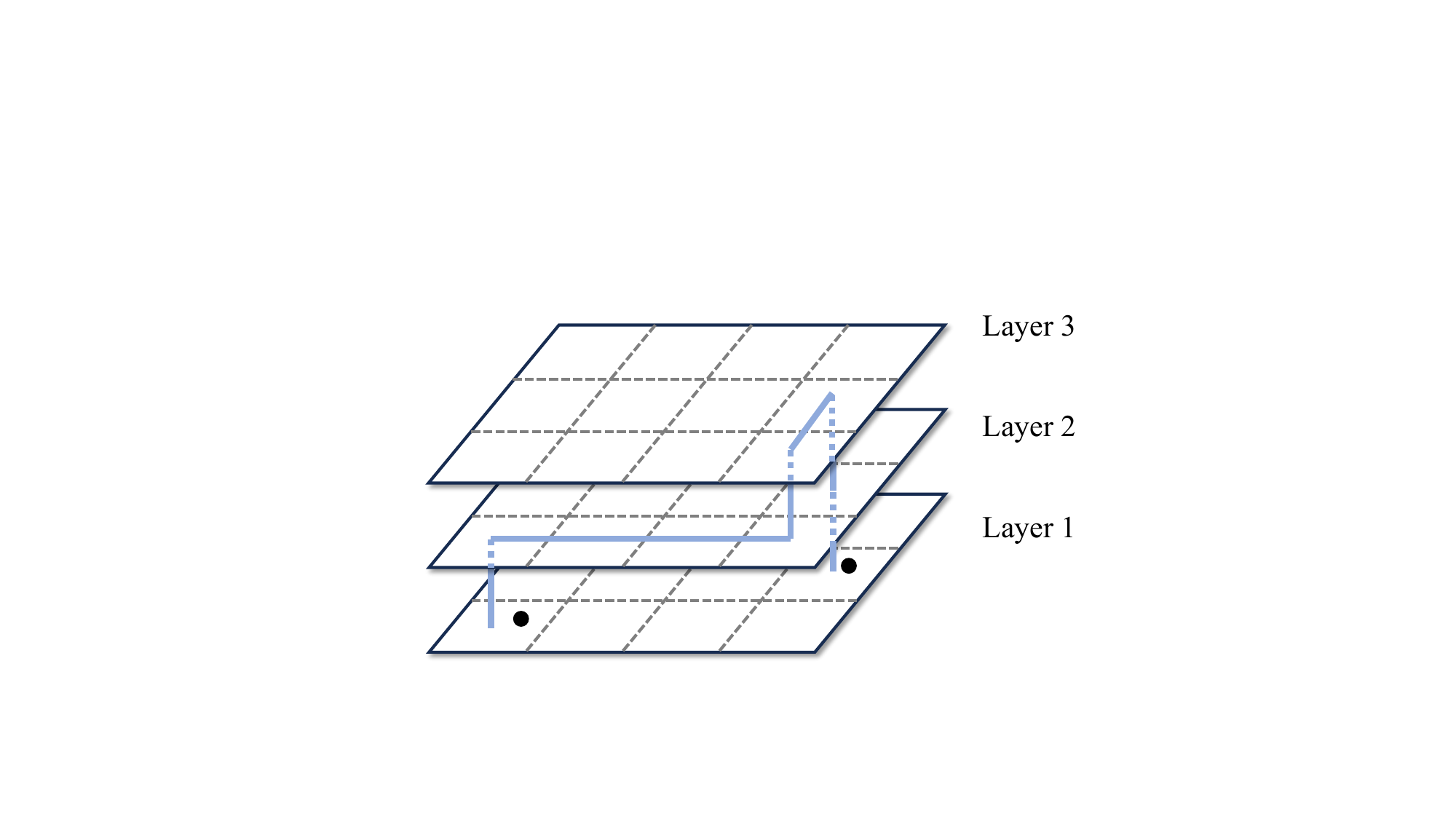}
        \label{fig:3d}
    }
    \caption{(a) 2D global routing solution for a net. (b) 3D global routing solution after layer assignment process.}
\end{figure}

\subsection{GCell in Global Routing}
In the modern physical design flow, as demonstrated in Figure~\ref{fig:GCell graph}, the 3D multi-layer design is partitioned into small rectangular units called GCells by (usually evenly spaced) horizontal and vertical grid lines.
To align with the actual manufacture, routing segments are constrained to  this grid graph and each edge is assigned a capacity value to characterize its available routing resource.
We denote the edge between two GCells on the same layer wire and the edge between two GCells on the adjacent layers via.
In a traditional global routing flow, the router finds a 2D solution and then applies layer assignment to get a well optimized 3D solution.

\begin{figure}[tb]
    \centering
    \includegraphics[width=\linewidth]{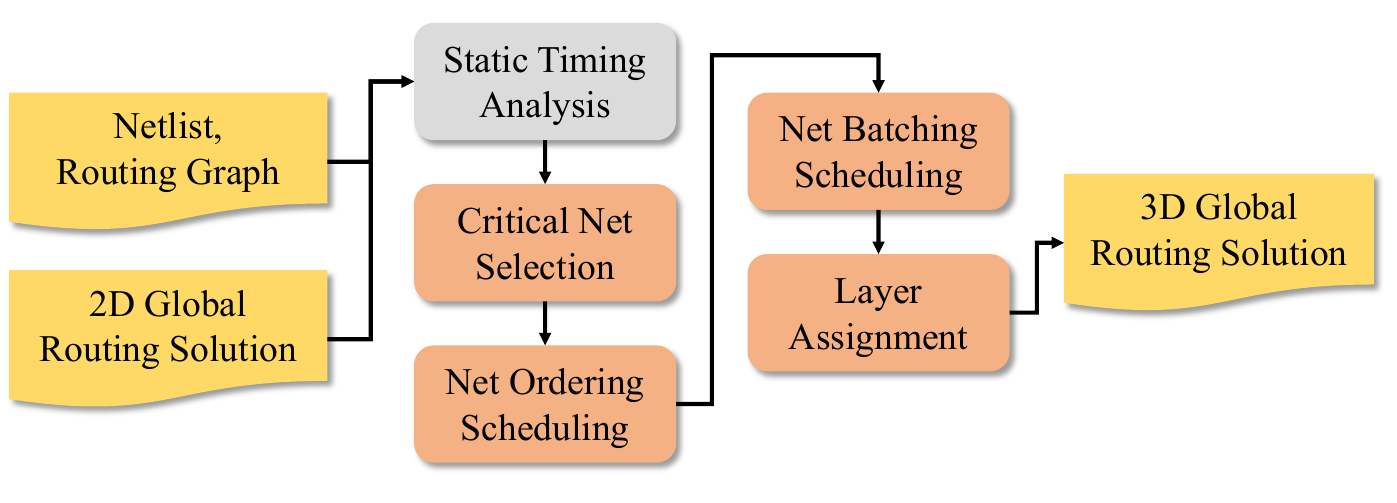}
    \caption{The overall framework of \texttt{GAP-LA}.}
    \label{fig:framework}
\end{figure}

\subsection{Layer Assignment}
Layer assignment typically takes a GCell grid graph with GCell edge capacity, a netlist and an optimized 2D global routing solution as input and assigns each 2D routing segment to a certain layer to generate a 3D global routing solution.
As illustrated in Figure~\ref{fig:2d} and Figure~\ref{fig:3d}, the layer assignment process successfully derives a 3D solution. Notably, this newly obtained 3D solution maintains the exact same projection as the input 2D solution.
A typical layer assignment flow reorders the input nets and performs tree dynamic programming kernel to optimize each net in sequential order.
The common objective is usually weighted sum of total overflow and via count.
In performance-driven layer assignment problem, the objective also involves worst negative slack (WNS), total negative slack (TNS) and power consumption, which are reported by static timing analysis and power analysis tool.

\subsection{Static Timing Analysis}
Static timing analysis (STA) is an essential step to ensure the performance of the design during the physical design flow.
It represents the circuit as a directed acyclic graphs (DAG), where nodes denote pins of circuit components and edges denote pin interconnects.
A typical STA engine performs timing propagation to obtain arrival time, required arrival time, slack, and eventually critical paths. 
After these two steps, slack of pins, critical paths and other timing information are reported.
The slack of a pin is the difference between its required arrival time and arrival time and describes the extent of timing violation of the pin.
The critical path of a primary output leads to the longest delay and worst slack from primary input to it.
Many different kinds of delay models can be used by an STA engine.
In our framework, we compute cell delays though the nonlinear delay model (NLDM) based on 2D lookup tables with load and slew indices and Elmore delay model~\cite{Elmore1948} is applied to get net delay.
These models are widely used in recent academic research and contests.

\subsection{Problem Formulation}
Given a layout partitioned into rectangular subregions (GCells) by a set of horizontal and vertical grid lines, and a GCell-level 2D routing solution for nets, preformance-driven layer assignment determines a 3D routing solution for each net. 
The process simultaneously optimizes WNS, TNS, power consumption, and total overflow.
Critically, the planar projection of the 3D result must strictly match the input 2D solution.

\subsection{Evaluation Metrics}
In this work, we follow the same evaluation method as the ISPD25 Performance-driven Large Scale Global Routing Contest~\cite{ISPD2025}, which is defined as follows:

\textbf{Quality score}: Quality score is defined by the weighted sum of WNS, TNS, power consumption and total overflow,
\begin{equation}
     w_1\cdot(WNS-r_1)+w_2\cdot\frac{TNS-r_2}{N_{end}}+w_3\cdot(pow-r_3)+w_4\cdot tof,
\end{equation}
where $w_1, w_2, w_3, w_4$ are input for each design to adjust the weights of different metrics and $r_1, r_2, r_3$ are input reference metric value.
$N_{end}$ is the number of end points in the design and $tof$ is total overflow of all the GCell edges.
The end points usually indicate the primary outputs and D-pins of flip-flops in the design.
In our evaluation flow, WNS, TNS and power are reported by open source tool OpenROAD~\cite{Ajayi2019OpenROAD}.
Note that, value of $r_1, r_2, r_3$ does not influence the comparison among different solutions because value of $w_1\cdot r_1, w_2\cdot r_2, w_3\cdot r_3$ is constant.

\textbf{Overflow}: For a GCell edge $e$ on layer $l$, its overflow is usually computed with its demand $d$ and capacity $c$ in previous works as follows,
\begin{equation}
    of=max\left\{0, d -c\right\}.\label{eq:relu}
\end{equation}
To align with ISPD25 Contest evaluation metrics~\cite{ISPD2025}, we use
\begin{equation}
    of = ofw\left(l\right)\cdot e^{s\left(d-c\right)}\label{eq:exp}
\end{equation}
as its overflow, where $s$ is $0.5$ for $c > 0$, $1.5$ for $c = 0$, and $ofw\left(l\right)$ denotes the overflow weight of layer $l$, which is given in the benchmark.
    The overflow metrics in Eq.~\ref{eq:exp} is better than that in Eq.~\ref{eq:relu}, because it penalizes a highly congested edge in the routing solution, which makes it very difficult to resolve design rule violations in the detailed routing stage.

\section{Algorithms}
\label{sec:algo}

\begin{figure}[t]
    \centering
    \includegraphics[width=0.90\linewidth]{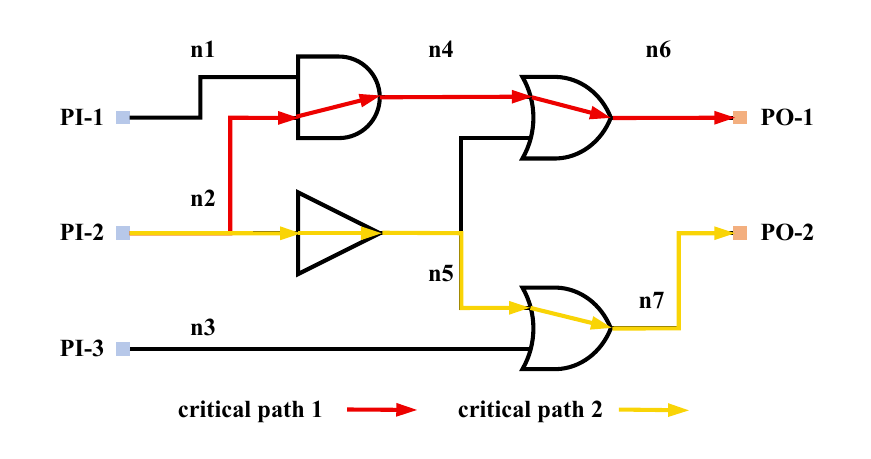}
    \caption{An example of critical paths in the static timing analysis graph.}
    \label{fig:critical}
\end{figure}

\begin{algorithm}[b]
    \caption{Critical Net Analysis-based Net Ordering and Batching Strategy.}\label{alg:order}
    \begin{algorithmic}[1]
        \REQUIRE netlist $N$, 2D routing solution $GRS$-$2D$, \\
        \ \quad average resistance of unit length metal $r_{avg}$, \\
        \ \quad average capacitance of unit length metal $c_{avg}$, \\
        \ \quad static timing analysis engine STA, semi-critical slack\\
        \ \quad ratio $\alpha$,threshold $th$ \\ 
        \ENSURE Batches $BS$\\
            \STATE $PinSlack, CritPaths$ $\leftarrow$ \texttt{STA}($GRS$-$2D$, $r_{avg}$, $c_{avg}$, $\alpha$) \\
            \STATE $C_{N}$ $\leftarrow$ \texttt{CountCritPaths}($N$, $CritPaths$) \\
            \STATE $N_{c}, N_{s}, N_{n}$$\leftarrow$\texttt{Divide}($N$, $C_{N}$, $PinSlack$, $th$)\\
            \STATE $\{S_{c, 0}, S_{c, 1}...\}\leftarrow$\texttt{Partition}$\left(S_c, C_N\right)$ \\
            \STATE $\{S_{s, 0}, S_{s, 1}, ...\}\leftarrow$\texttt{Partition}$\left(S_n, PinSlack\right)$
            \STATE $BS_{c}\leftarrow$\texttt{GetBatch}$\left(\{S_{c, 0}, S_{c, 1}, ...\}\right)$ \\
            \STATE $BS_{s}\leftarrow$\texttt{GetBatch}$\left(\{S_{s, 0}, S_{s, 1}, ...\}\right)$\\
            \STATE $BS_{n}\leftarrow$\texttt{GetBatch}$\left(S_n\right)$ \\
        \STATE $BS$ $\leftarrow$ \texttt{Concat}($BS_{c}, BS_{s}, BS_{n}$)
    \end{algorithmic}
\end{algorithm}

Figure~\ref {fig:framework} illustrates the overall framework of \texttt {GAP-LA}, which takes a netlist, routing graph, and 2D global routing solution as inputs and ultimately generates a well-optimized 3D global routing solution. 
After loading the netlist, global routing graph, and 2D solution data, we conduct a static timing analysis on the GPU using \texttt {HeteroSTA}~\cite {gputimertcad23, heteroexcepticcad24, HeteroSTAWebsite}.
Leveraging the timing report, we identify critical nets and prioritize their assignment.
Guided by net criticality insights, we carefully implement net ordering to enhance both performance and overflow outcomes. 
Next, we generate multiple net batches based on this ordering and perform parallel layer assignment for nets within the same batch on the GPU. 
Finally, our framework yields a 3D global routing solution with optimized performance metrics.

\subsection{Critical Net Analysis-based Net Ordering and Batching Strategy\label{subsec:order}}
The delay of critical paths is of paramount importance for design timing closure, and our solution aims to minimize critical path delay as much as possible. 
However, since existing work often adopts net-based~\cite{Lee08LayerAssign, Lee09, Xu09FastRoute4.0, Liu11NegoLayerAssign, Zhang20MiniDelay} or edge-based~\cite{Dai12NCTU-GR} optimization strategies for layer assignment, the delay of critical paths cannot be optimized directly. 
Therefore, identifying the criticality of each net is essential for our net-based approach.

Figure~\ref{fig:critical} depicts an example of a static timing analysis graph with three primary inputs and two primary outputs. 
    Our static timing analysis engine reports one critical path for each endpoint.
Critical path 1 traverses nets $n_2, n_4$, and $n_6$ while critical path 2 traverses nets $n_2, n_5$, and $n_7$.
We define net criticality as the number of critical paths traversing the net.
In example shown in Figure~\ref{fig:critical}, net $n_2$ exhibits the highest criticality, whereas nets $n_4, n_5, n_6$ and $n_7$ have lower criticality, and all other nets have the lowest criticality.
Intuitively, optimizing a net with higher criticality yields greater benefits, as it may dominate the arrival times of multiple primary outputs.


\begin{figure}[t]
    \centering
    \subfloat[]{
        \includegraphics[width=0.40\linewidth]{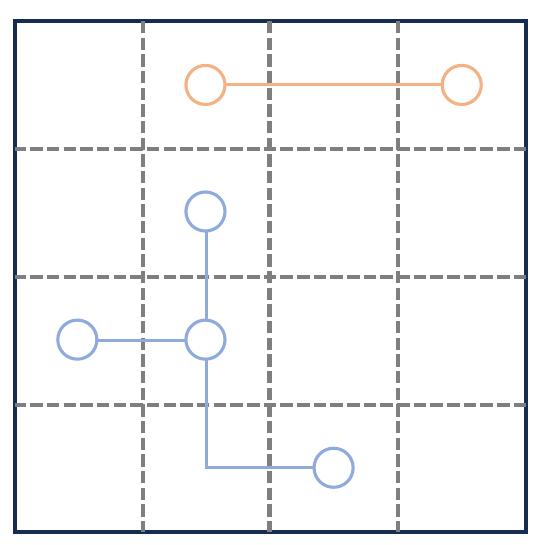}
        \label{fig:graph_a}
    }
    \subfloat[]{
        \includegraphics[width=0.40\linewidth]{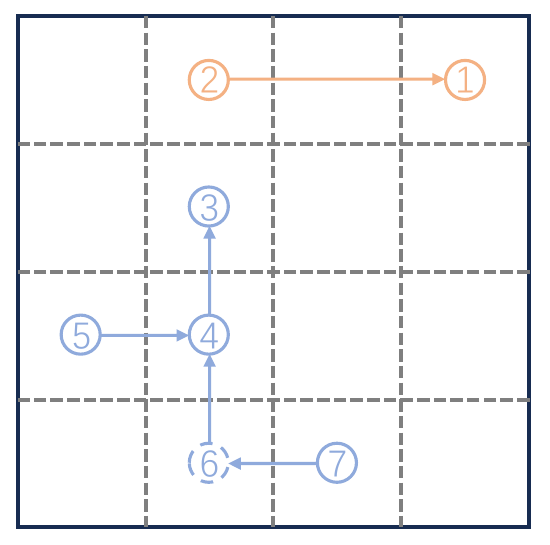}
        \label{fig:graph_b}
    }
    \caption{(a) The original 2D routing structure. Net 1 has 2 nodes and net 2 has 4 nodes.
             (b) Our layer assignment directed tree. Net 1 has 2 nodes and net 2 has 5 nodes.}
\end{figure}
\begin{figure}[b]
    \centering
    \subfloat[]{
        \label{fig:weighting}
        \includegraphics[width = 0.5\linewidth]{
            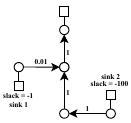
        }
    }
    \subfloat[]{
        \label{fig:cost}
        \includegraphics[width = 0.45\linewidth]{
            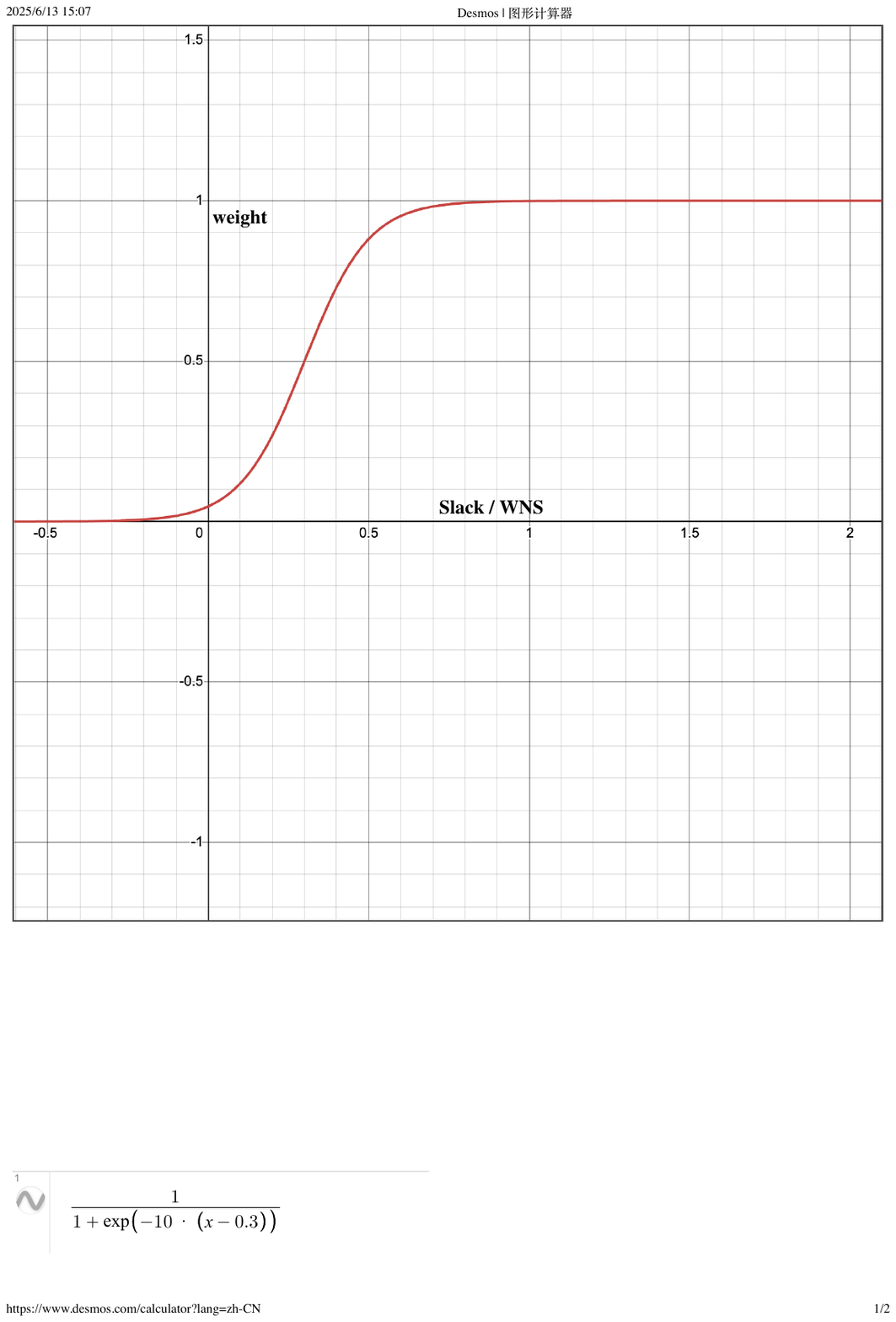
        }
    }
    \caption{(a) Direct tree edge weighting. (b) Pin weight is the logistic function of its $\frac{slack}{WNS}$.}
\end{figure}
\begin{figure}[t]
    \centering
    \includegraphics[width=0.7\linewidth]{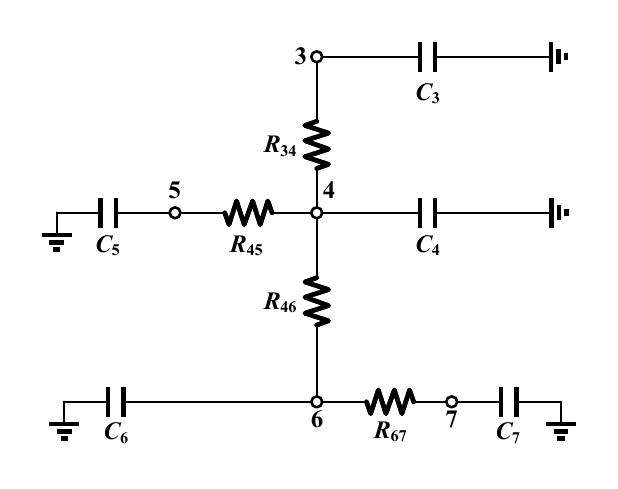}
    \caption{RC tree of net 2 in Figure~\ref{fig:graph_b}.}\label{fig:rc}
\end{figure}

In a typical sequential layer assignment flow, net ordering is a crucial determinant of the final routing solution quality.
We prioritize the critical nets identified earlier, as they exert the most significant impact on the arrival times of primary outputs.
In our flow, the static timing analysis engine reports the top-1 worst critical paths of every primary output with slack worse than $0.7\times$WNS.
Nets traversed by more than three relevant critical paths are designated as critical nets.
For each net, we define its net slack as the minimum slack value among its pins.
After processing these critical nets, nets with critical net slack(worse than $0.7\times$WNS) become critical, as low slack indicates they may dominate critical paths if the delay of original critical paths is reduced by assigning relevant nets to upper layers.
We call these nets semi-critical nets.
As an illustrated in Figure~\ref{fig:critical}, after optimizing critical nets $n_2, n_4, n_5, n_6$ and $n_7$, net $n_3$ may dominate the arrival time of PO-2, so that it should be prioritized.
Layer assignment for the remaining nets follows subsequently.

\begin{algorithm}[b]
    \caption{Layer Assignment\label{alg:layer assignment}}
    \begin{algorithmic}[1]
        \REQUIRE delay/capacitance/congestion weight $w^d$, $w^{cap}$ and $w^{cong}$, batches $BS$,
                    node son/parent node/pin  list $sons, par, pin$,  node position $x, y$, resistance (R) and capacitance (C) of unit length on each layer $r, c$, 
                        R of via on each layer $vr$,  estimated upstream R of each node $ur$,
                        downstream load C of candidate solution $dlc$,
                        the lowest layer with pin of a node $nl$, the highest layer with pin of a node $nh$, layout size $X, Y, L$, 
                        demand map $D$ and capacity map $C$\\
        \ENSURE 3D routing solution $GRS$-$3D$ \\
        \FOR{$batch$ in $BS$}
            \FOR{$level$ = 0 \TO max level of $batch$}
                \STATE Call \texttt{getSubtreeCandidate}
            \ENDFOR
            \FOR{$level$ = max level of $batch$ \TO 0}
                \STATE Call \texttt{traceBackSolution}
            \ENDFOR
        \ENDFOR \\
    \end{algorithmic}
\end{algorithm}

However, the strategy described above is purely sequential and must be adapted for parallel computing.
Our observations indicate that nets with similar criticality and slack have comparable impacts on final design performance. 
As a result, strict net ordering should be relaxed: we partition the net set into multiple subsets, enabling parallel assignment within each subset with minimal quality degradation.
For critical and semi-critical nets, let $C$ denote the maximum criticality.
We place nets with criticality $\left[C, C\right]$ into the first subset, $\left[\frac{C}{2}, C\right)$ into the second subsets, $\left[\frac{C}{4}, \frac{C}{2}\right)$ into the third subsets and so on.
For other nets with bad slack, we place nets with slack $\left[WNS, WNS\right]$ into the first subset, $\left[0.99\cdot WNS, WNS\right)$ into the second subset, $\left[0.96\cdot WNS, 0.99\cdot WNS\right)$ into the third subset and so on.
For non-critical nets (better than $0.7\times$WNS), we employ a traditional congestion-driven strategy instead, as these nets are far less likely to degrade design performance. 
Algorithm~\ref{alg:order} outlines our critical net analysis-based net ordering and batching strategy flow.
    We get pin slack and critical paths from timing report offered by a static timing analysis engine and count the number of critical paths passing through each net in Line 1 and Line 2.
    Then the netlist $N$ is divided into three subsets, $N_c, N_s, N_n$ according to the timing information in Line 3. 
    For timing-critical nets in $N_c, N_n$, we partition them with the strategy described above to prioritize more critical nets in Line 4 and Line 5. 
    In Line 6-8, we implement classical batching method used in~\cite{Liu20CUGR, Zhao25helemgr} to generate batches.

\renewcommand{\algorithmicrequire}{\textbf{Input:}}
\renewcommand{\algorithmicensure}{\textbf{Output:}}

\subsection{Layer Assignment Directed Tree}

To facilitate the extraction of parasitic parameters and the implementation of a fine-grained parallel layer assignment algorithm, we designed a directed tree structure.
In our layer assignment algorithm, we assume that a piece of 2D routing segment with no bend points should be assigned to a single layer without layer shifting, which is only performed at bends and Steiner points.
To satisfy these constraints, we introduce additional nodes to ensure all inter-node connections have no bend points, so layer shifting operations are confined to nodes in the directed tree.
As illustrated in Figure~\ref {fig:graph_a} and Figure~\ref {fig:graph_b}, node 6 is added to bridge node 4 and node 7, ensuring no routing paths with bend points exist in the directed tree.

Based on the tree discussed above, we extract parasitic parameters for static timing analysis.
The directed tree acts as a subgraph in the STA graph, with nodes connected to relevant pin nodes in the STA graph via zero-resistance edges.
For horizontal wire connections, we calculate resistance  and capacitance using the average per-unit-length resistance and capacitance of all horizontal wires. 
For vertical wire connections, we follow the same approach.
We then compute the resistance of all edges and capacitance of all nodes in the STA graph using the widely adopted $\pi$-model.

\begin{figure}[t]
    \centering
    \includegraphics[width = 0.8\linewidth]{
        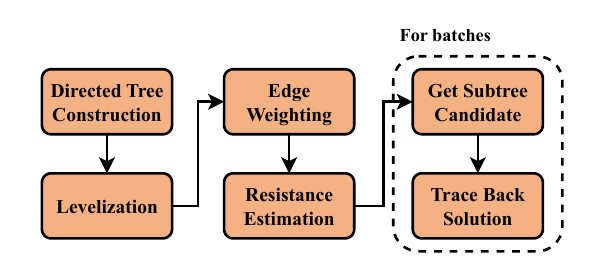
    }
    \caption{Layer assignment flow.\label{fig:assign}}
\end{figure}
\begin{algorithm}[b]
    \caption{Get Subtree Candidate\label{alg:get subtree candidate}}
    \begin{algorithmic}[1]
        \REQUIRE $level$ to get candidate \\
        \ENSURE candidate solution cost $f$, candidate solution look ahead cost $f'$, candiate solution layer selection $entry$ \\
        \texttt{Kernel 1:} \\
        \STATE $idx\leftarrow$threadIdx.x + blockDim.x $\times$ blockIdx.x \\
        \STATE $n, b, l\leftarrow$\texttt{GetTaskId}$\left(level, idx\right)$\\
        \FOR{$t$ = b \TO $L - 1$}
            \STATE $f_{n, b, t, l}\leftarrow$\texttt{GetCost}$\left(sons_n, b, t, l\right)$\\
            \STATE $f'_{n, b, t, l}, entry_{n, b, t, l}\leftarrow$ \\
            \texttt{GetLookAheadCost}$\left(sons_n, b, t, l\right)$\\
        \ENDFOR \\
        \texttt{Kernel 2:}\\
        \STATE $idx\leftarrow$threadIdx.x + blockDim.x $\times$ blockIdx.x \\
        \STATE $n, l\leftarrow$\texttt{GetTaskId}$\left(level, idx\right)$\\
        \STATE $b', t'\leftarrow\arg\min_{l\in \left[b, t\right]}f'_{n, b, t, l}$ \\
        \STATE $f_{n, l}, f'_{n, l}, entry_{n, l}\leftarrow f_{n, b', t', l}, f'_{n, b', t', l}, entry_{n, b', t', l}$\\
    \end{algorithmic}
\end{algorithm}

\subsection{Directed Tree Edge Weighting}
Due to the fact that each net has distinct criticality, our algorithm allocates varying optimization efforts to reduce the delay and capacitance of different nets. 
In prior studies such as ~\cite{Liao22DreamPlace4.0}, net weighting has emerged as a prevalent approach to tackle this challenge.

However, this approach overlooks the fact that each sink of a net has distinct criticality. 
As demonstrated in Figure~\ref{fig:weighting}, sink 1 has estimated slack of -1 ps and sink 2 has estimated slack of -100 ps.
Since sink 2 exhibits a significantly worse negative slack than sink 1, optimizing the delay from the net source to sink 2 warrants greater effort.
Figure~\ref{fig:cost} illustrates that the pin weight follows a logistic function of $\frac{slack}{WNS}$:
\begin{equation}
    w_{pin} = \frac{1}{1 + k\left(\frac{slack}{WNS} - b\right)},
\end{equation}
 and we select $k=10, b = 0.3$ in our implementation.
 \begin{algorithm}[t]
    \caption{Trace Back Solution\label{alg:trace back solution}}
    \begin{algorithmic}[1]
        \REQUIRE cost list $f$, choice list $choice$, $level$ to trace back \\
        \ENSURE 3D routing solution $GRS$-$3D$ \\
        \STATE $idx\leftarrow$ threadIdx.x + blockDim.x $\times$ blockIdx.x \\
        \STATE $n\leftarrow$\texttt{GetLevelNode}$\left(level, idx\right)$ \\
        \IF{$n$ is root node}
            \STATE $best_n\leftarrow\arg\min_{l\in \{0, 1, \cdots, L - 1\}}f_{n, l}$ \\
        \ENDIF
        \STATE $b, t\leftarrow choice_{n, best}$ \\
        \STATE \texttt{Add via }$\left(x_n, y_n, b\right)$-$\left(x_n, y_n, t\right)$\texttt{ to }$GRS$-$3D$\\
        \FOR{$son$ \textbf{in} $sons_n$}
            \STATE $best_{son}\leftarrow entry_{n, best_n, son}$\\
            \STATE \texttt{Add wire}\\
            $\left(x_n, y_n, best_{son}\right)$-$\left(x_{son}, y_{son}, best_{son}\right)$\texttt{ to }$GRS$-$3D$ \\
        \ENDFOR
    \end{algorithmic}
\end{algorithm}

The logistic function value remains nearly 1 as $\frac{slack}{WNS}$ approaches 1, enabling our algorithm to effectively optimize paths with poor slack.
For an edge in the directed tree, its weight is defined as the maximum weight of all sinks in its downstream subtree:
\begin{equation}
    w_{e:u\rightarrow v} = \max_{pin\in subtree\left(u\right)}{w_{pin}}
    \label{eq:edge weight}
\end{equation}


\subsection{Layer Assignment with Look Ahead Strategy}
To concurrently optimize slack, power consumption, and congestion, we integrate three cost functions: net delay cost, net capacitance cost, and net congestion cost.
Net capacitance is prioritized here, as it directly influences both power consumption and cell delay. 
In the Nangate 45nm library, cell delay dominates path delay: a cell may exhibit equivalent drive resistances exceeding 1k$\Omega$, whereas the resistance of a GCell edge remains below 0.01k$\Omega$. 
In a typical congestion- and via count-driven sequential layer assignment framework, we employ tree dynamic programming (DP) to optimize the objective for each net. The tree DP method proceeds bottom-up to compute the optimal solution for each candidate subtree, then top-down to backtrack the optimal solution for the entire tree. 
For each node, DP retains the optimal solution for each possible layer through which the path from the node to its parent traverses.
However, this method is ineffective for net delay optimization due to the after-effect.

\begin{table}[t]
    \centering
    \caption{Details of Modified NanGate 45nm library Used in the Contest.\label{table:Nangate45}}
    \resizebox{\linewidth}{!}{
    \begin{tabular}{|c|ccc|}
    \hline
        Layer & R (k$\Omega$/$\mu$m) & C (fF/$\mu$m) & R of Upper Via  (k$\Omega$) \\
    \hline 
        metal1 & $5.4286 \times 10^{-3}$ & $7.41819 \times 10^{-2}$ & $5\times 10^{-3}$\\
        metal2 & $3.5714 \times 10^{-3}$ & $6.74606 \times 10^{-2}$ & $5\times 10^{-3}$\\
        metal3 & $3.5714 \times 10^{-3}$ & $8.88758 \times 10^{-2}$& $5\times 10^{-3}$\\
        metal4 & $1.5000 \times 10^{-3}$ & $1.07121 \times 10^{-1}$&$3\times 10^{-3}$ \\
        metal5 & $1.5000 \times 10^{-3}$ & $1.08964 \times 10^{-1}$& $3\times 10^{-3}$\\
        metal6 & $1.5000 \times 10^{-3}$ & $1.02044 \times 10^{-1}$ &$3\times 10^{-3}$\\
        metal7 & $1.8750 \times 10^{-4}$ & $1.10436 \times 10^{-1}$ & $1\times 10^{-3}$\\
        metal8 & $1.8750\times 10^{-4}$ & $9.69714 \times 10^{-2}$& $1\times 10^{-3}$\\
        metal9 & $3.7500 \times 10^{-5}$ & $3.6864 \times 10^{-2}$&$0.5\times 10^{-3}$\\
        metal10 & $3.7500 \times 10^{-5}$ & $2.8042 \times 10^{-2}$ & -\\
    \hline
    \end{tabular}
    }
\end{table}

Figure~\ref{fig:rc} shows the RC tree extracted from net 2 in Figure~\ref{fig:graph_b} using the $\pi$-model, where node 3 is the root of the tree and node 5, 6, 7 are the leaves.
In the Elmore delay model, the delay from node 3 to node 4 $d_{34}$  is computed as $R_{34} \times \left(C_4 + C_5 + C_6 + C_7\right)$.
The subtree choices for nodes 5, 6, and 7 introduce after-effects on $d_{34}$, which is absent in congestion or wirelength cost calculations. 
\begin{table*}[b]
    \centering
    \caption{Benchmark details.}
    \label{table:details}
    \begin{threeparttable}
        
    \begin{tabular}{|c|c|c|c|c|c|c|}
        \hline
        Case$^\dag$ & Name & $w_{1}/w_{2}/w_{3}/w_{4}$ & $r_1/r_2/r_3$ & $N_{end}$ & $N_{net}$  & GCell Graph Dimensions\\
        \hline
        1 & ariane\_v & 10 / 100 / 300 / 3e-6 & -0.485 / -1398.39 / 0.646 & 20218 & 123900 & 10$\times$761$\times$761\\
        2 & bsg\_v & -10 / -100 / 25 / 4e-8 & -0.44 / -10802.7 / 3.05 & 214821 & 736883 & 10$\times$1384$\times$1384 \\
        3 & nvdla\_v & -0.05 / -0.5 / 25 / 1.5e-7 & -94.78 / -669471 / 2.96 & 45925 & 199481 & 10$\times$1120$\times$1120 \\
        4 & tile\_v & -1 / -10 / 300 / 7e-7 & -0.695 / -3590.83 / 0.1455 & 13350 & 136120 & 10$\times$428$\times$428 \\
        5 & group\_v & -1 / -10 / 20 / 3e-8 & -0.815 / -41740.2 / 7.82 &347869 & 3274611 & 10$\times$1611$\times$1610 \\
        6 & cluster\_v & -1 / -10 / 0.3 / 5e-9 & -0.68 / -79748 / 23.7 & 1082397 & 12047279 & 10$\times$3175$\times$3175 \\
        7 & ariane\_b & -0.2 / -2 / 100 / 4e-7 & -1.628 / -523.038 / 0.156  & 20218 & 105924 &  10$\times$646$\times$646\\
        8 & bsg\_b& -0.1 / -1 / 50 / 2e-8 & 0 / 0 / 0.305 & 214915 & 768239 & 10$\times$1384$\times$1384  \\
        9 & nvdla\_b & -0.01 / -0.1 / 100 / 1e-7 & 0 / 0 / 0.136 & 45925 & 157744 &10$\times$1120$\times$1120  \\
        10 &tile\_b & -3 / -30 / 100 / 1e-6 & -0.483 / -1920.724 / 0.145 & 13350 & 135814 & 10$\times$386$\times$386 \\
        11 & group\_b & -0.5 / -5 / 3 / 4e-8 & -0.68 / -40487.883 / 8.547 & 347869 & 3218496 & 10$\times$1611$\times$1610 \\
        12 & cluster\_b& -0.4 / -4 / 2 / 1e-8 & -0.29 / -53249 / 24.261 & 1082397 & 12168735 &10$\times$3719$\times$3719 \\
        \hline
    \end{tabular}
        \begin{tablenotes}
            \footnotesize
            \item $\dag$ *\_v and *\_b denote the visible and blind cases in the contest, respectively.
        \end{tablenotes}
    \end{threeparttable}
\end{table*}

Due to the significant resistance of vias, routing a simple 2-pin net on lower layers reduces net delay but increases net capacitance. 
As a result, small subtrees tend to favor lower layers for reduced delay, which can degrade the delay of upstream connections.
To address this, we propose a look-ahead strategy. 
Let $f_u$ denote the cost of the subtree rooted at node $u$ comprising delay, capacitance, and congestion costs.
In our layer assignment directed tree, we estimate the upstream resistance of each node and use $f'_u=f_u+w^{d}\times r_u^{est} \times c_u$ to select optimal candidate instead of $f_u$, where $w^d$ is the weight of delay cost, $r^{est}_u$ is the estimated upstream resistance of node $u$ and $c_u$ is the downstream capacitance of node $u$.

This approach anticipates after-effects by evaluating how downstream capacitance contributes to the entire tree’s delay cost.

For example in Figure~\ref{fig:graph_b} and ~\ref{fig:rc}, we first estimate resistances $R^{est}_{34}, R^{est}_{45}, R^{est}_{46}$ and $R^{est}_{67}$ using node distances and the average unit-length metal resistance  $r_{avg}$.
We then use $f'_6 = w^{cong} \times cong_{67} + w^{cap}\times c_{67} + w^{d}\times (d_{67} + r^{est}_{6}\times c_{6})$ to evaluate all the candidates solution for subtree 6.

Figure~\ref{fig:assign} illustrates our layer assignment flow.
After the preparation steps, we obtain the directed tree structure, level information, weighting of each edge, and resistance values for the look-ahead strategy. 
The algorithm then proceeds to the most computation-intensive part in Algorithm~\ref{alg:layer assignment}.

Algorithm~\ref{alg:get subtree candidate} details our GPU kernel implementation.
We design two kernels to calculate the node cost, node look-ahead cost and layers used by children nodes considering a node connects to its parent node via different layers.
The first kernel is used to enumerate all possible via connection pairs $\left(b, t\right)$ connect layer $b$ to layer $t$) used at node $n$ and its son nodes when the node $n$ connects its parent node via layer $l$ (Line 3-6).
The optimal legal candidate with minimal look-ahead cost is chosen for each son.
The reason why we introduce layer $l$ is that layer $l$ determines the via delay to its children nodes.
After the enumeration, we use the second kernel to calculate the best solution based on data calculated in the first kernel.

Algorithm~\ref{alg:trace back solution} obtains the 3D routing solution from the candidate subtrees computed in Algorithm~\ref{alg:get subtree candidate}.
The root node solution with minimal total cost is selected, and the corresponding subtree solutions for its sons are backtracked. 
Though conceptually recursive, the implementation uses a GPU-friendly BFS process.
\section{Experimental Results}
\label{sec:exp}

\begin{figure*}
\begin{minipage}{\linewidth}
    \centering
    \captionof{table}{Comparison of quality score, runtime and via count with top-3 winners in the ISPD 2025 contest.}
    \label{table:QoR and RT}
    \resizebox{\linewidth}{!}{
        \begin{tabular}{|c|ccccccc|cccc|cc|cccc|}
        \hline
        & \multicolumn{7}{c|}{\textbf{Quality Score}} & \multicolumn{4}{c|}{\textbf{Runtime(s)}} & \multicolumn{2}{c|}{\textbf{GAP Runtime(s)}} &
        \multicolumn{4}{c|}{\textbf{Via Count}}\\
        & 1st & Imp. & 2nd & Imp. & 3rd & Imp. & Ours & 1st & 2nd & 3rd & Ours & GPU & CPU & 1st & 2nd & 3rd & Ours \\
        \hline
        case1 & 0.896 & -2.055 & 0.639 & -1.797 & 0.094 & -1.253 & \textbf{-1.158} & \textbf{6} & \textbf{6} & 15 & 8 & 2 & 2 & 869552 & \textbf{797909} & 1452629 & 1040032 \\
        case2 & 0.562 & -0.640 & 0.722 & -0.800 & 0.094 & -0.172 & \textbf{-0.078} & 28 & \textbf{21} & 30 & 27 & 6 & 11 & 5100730 & \textbf{4575171} & 6483299 & 5422990 \\
        case3 & -1.995 & -0.137 & -2.077 & -0.055 & -1.824 & -0.308 & \textbf{-2.132} & 8 & \textbf{6} & 17 & 9 & 4 & 3 & 1393389 & \textbf{1270319}& 2147099 & 1384821 \\
        case4 & \textbf{-0.405} & +0.072 & 0.401 & -0.734 & 1.370 & -1.703 & -0.333 & \textbf{7} & 8 & 16 & 6 & 2 & 3 & 1181943 & \textbf{1014414} & 1376740 & 1062075 \\
        case5 & -2.029 & -1.671 & -1.295 & -2.406 & -3.599 & -0.101 & \textbf{-3.700} & \textbf{115} & 134 & 192 & 113 & 24 & 46 & 23842158 & \textbf{21381686} & 32437038 & 26670038 \\
        case6 & 0.510 & -0.135 & 0.399 & -0.024 & 0.714 & -0.339 & \textbf{0.375} & \textbf{388} & 476 & 687 & 421 & 58 & 139 & 89575117 & \textbf{75139223} & 82079269 & 91771657 \\
        case7 & 1.792 & -0.267 & 1.659 & -0.133 & 1.531 & -0.005 & \textbf{1.525} & \textbf{7} & \textbf{7} & 16 & \textbf{7} & 3 & 3 & 830056 & 893782 & \textbf{765229} & 810382 \\
        case8 & 0.849 & -0.409 & \textbf{0.415} & +0.026 & 0.689 & -0.248 & 0.441 & 30 & \textbf{16} & 38 & 31 & 8 & 10 & 5404700 & 4761540 & \textbf{4704104} & 5182861 \\
        case9 & 1.431 & -0.077 & 1.661 & -0.307 & 1.392 & -0.038 & \textbf{1.354} & 8 & \textbf{6} & 18 & 8 & 3 & 3 & 1319759 & 1172952 & \textbf{1136300}& 1209985 \\
        case10 & 0.344 & -0.122 & 2.074 & -1.853 & 1.803 & -1.582 & \textbf{0.221} & \textbf{7} & 8 & 16 & 6 & 2 & 3 & 1060382 & \textbf{967350} & 1079131 & 992786 \\
        case11 & 1.332 & -0.000 & \textbf{1.232} & +0.099 & 3.116 & -1.784 & 1.331 & 114 & \textbf{96} & 173 & 106 & 25 & 43 & 23292577 & \textbf{20969891} & 33044068 & 26158463 \\
        case12 & 2.327 & -0.500 & 2.370 & -0.544 & 2.677 & -0.851 & \textbf{1.826} & 391 & \textbf{385} & 655 & 431 & 60 & 156 & 89609278 & \textbf{79136523} & 116361040 & 91495810 \\
        \hline
        \hline
        Avg. & 0.468 & 0.495 & 0.683 & 0.711 & 0.671 & 0.699  & \textbf{-0.027} & \textbf{92} & 97 & 156 & 98 & 16 & 35 & 20289970 & \textbf{17673397} & 23588829 & 21100158 \\
        \hline
        \end{tabular}
    }
    \captionof{table}{Comparison of WNS, TNS, power and congestion with top-3 winners in the ISPD 2025 contest.}
    \vspace{0.2cm}
    \label{table:metrics}
    \resizebox{\linewidth}{!}{
        \begin{tabular}{|c|cccc|cccc|cccc|cccc|}
            \hline 
            & \multicolumn{4}{c|}{\textbf{WNS}} & \multicolumn{4}{c|}{\textbf{TNS}} & \multicolumn{4}{c|}{\textbf{Power}} & \multicolumn{4}{c|}{\textbf{Congestion}} \\
            & 1st & 2nd & 3rd & Ours & 1st & 2nd & 3rd & Ours & 1st & 2nd & 3rd & Ours & 1st & 2nd & 3rd & Ours \\
            \hline
            case1 & -0.432 & -0.398 & -0.381 & {-0.365} & -1325.8 & -1340.2 & -1138.4 & {-997.9} & 0.646092 & 0.646078 & {0.646080} & 0.646076 & {5846461} & 5902619 & 8003814 & 6662339 \\
            case2 & -0.420 & -0.424 & -0.407 & {0.388} & -10366.2 & -10774.9 & {-9344.3} & -9602.9 & 3.05441 & {3.05301} & 3.05345 & 3.05349 & 21332547 & {20468992} & 25412296 & 22947622 \\
            case3 & -56.919 & {-55.841} & -55.914 & -56.253 & -521344.6 & {-514660.3} & -514674.3 & -515014.3 & 2.94054 & 2.94011 & 2.93806 & {2.93766} & {13315606} & 13686714 & 15687876 & 13562374 \\
            case4 & {-0.379} & -0.489 & -0.520 & -0.387 & {-1610.0} & -2225.5 & -2524.3 & -1711.6 & {0.14310} & 0.14515 & 0.14522 & 0.14383 & 3019425 & {2478318} & 3469019 & 2689765 \\
            case5 & -0.305 & -0.347 & {-0.275} & -0.327 & -20473.8 & -21709.1 & -19338.7 & {-17619.7} & 7.69208 & 7.73386 & {7.55296} & 7.59214 & 55005746 & {49071540} & 97506553 & 67934233 \\
            case6 & -0.266 & -0.315 & -0.276 & {-0.247} & -59712.5 & -62460.0 & -59533.5 &  {-52800.0} & 23.4497 & 23.5781 & 23.2200 & {23.1486} & 236920427 & {192012708} & 289712328 & 24407995 \\
            case7 & -1.756 & -0.737 & {-0.633} & -0.645 & -650.1 & -149.5 & -162.4 & {-111.5} & 0.15614 & 0.15613 & {0.15611} & 0.15614 & {4349151} & 4651055 & 4386728 & 4371048 \\
            case8 & -3.984 & {0.000} & -2.747 & {0.000} & -2859.8 & {0.000} & -2936.0 & {0.000} & {0.30473} & 0.30477 & 0.30478 & 0.30481 & 22567636 & 21307154 & {20564539} & 22506663 \\
            case9 & {0.000} & {0.000} & {0.000} & {0.000} & {0.000} & {0.000} & {0.000} & {0.000} & {0.13625} & 0.13626 & 0.13628 & 0.13629 & 14061155 & 16352692 & 13637302 & {13242873} \\
            case10 & {-0.342} & -0.472 & -0.440 & 0.353 & -1356.1 & -1920.7 & -1806.8 & {-1312.2} & {0.14303} & 0.14485 & 0.14520 & 0.14390 & 2232008 & 2122543 & 2168599 & {2087873} \\
            case11 & -0.433 & -0.460 & {-0.373} & -0.404 & -32016.1 & -34846.1 & {-26801.4} & -27116.0 & 8.36880 & 8.35859 & {7.94192} & 8.19256 & 52783071 & {49722314} & 132027917 & 68117588 \\
            case12 & -0.223 & -0.294 & -0.200 & {-0.199} & -42564.1 & -50372.4 & -39516.2 & {-38106.6} & 24.2439 & 24.3801 & {23.9679} & 24.0595 & 242737068 & {214110807} & 335020596 & 232170935 \\
            \hline
            \hline
            Avg. & -5.455 & -4.981 & -5.181 & \textbf{-4.964} & -57856.6 & -58371.6 & -56481.4 & \textbf{-55366.1} & 5.93989 & 5.96475 & \textbf{5.85066} & 5.87625 & 56180858 & \textbf{49323955} & 78966464 & 58391776 \\
            Ratio & 1.099 & 1.003 & 1.044 & \textbf{1.000} & 1.045 & 1.054 & 1.020 & \textbf{1.000} & 1.011 & 1.015 & \textbf{0.996} & 1.000 & 0.962 & \textbf{0.845} & 1.352 & 1.000 \\
            \hline
        \end{tabular}
    }
\end{minipage}
\end{figure*}

    

\begin{figure*}[tb]
\begin{minipage}{\linewidth}


    \centering
    \captionof{table}{Comparison of Overall Quality Scores, WNS, TNS, and power by replacing the layer assignment algorithms of top-3 winners and~\cite{Zhao25helemgr} with our GAP-LA .\label{table:replace}}
    \resizebox{0.55\linewidth}{!}{
    \begin{tabular}{|c|cccc|}
    \hline
         & \multicolumn{4}{c|}{\textbf{Quality Score}} \\
         & 1st + GAP-LA$^\dag$ & 2nd + GAP-LA$^\dag$ & 3rd + GAP-LA$^\dag$ & \cite{Zhao25helemgr} + GAP-LA$^\dag$ \\
         \hline
         case1 & -1.161 (+2.058) & -1.223 (+1.863) & -1.182 (+1.276) & -1.189 (+3.829) \\
         case2 & -0.054 (+0.616) & -0.127 (+0.849) & -0.196 (+0.290) & 0.013 (+1.016) \\
         case3 & -1.996 (+0.001) & -1.965 (-0.112) & -2.023 (+0.199) & -2.032 (+9.747) \\
         case4 & -0.369 (-0.036) & -0.453 (+0.854) & -0.435 (+1.805) & -0.319 (+1.805) \\
         case5 & -3.808 (+1.779) & -6.167 (+4.872) & -4.456 (+0.857) & -6.015 (+6.555) \\
         case6 & 0.384 (+0.126) & 0.366 (+0.033) & 0.329 (+0.385) & 0.397 (+0.711) \\
         case7 & 1.532 (+0.260) & 1.535 (+0.124) & 1.524 (+0.007) & 1.551 (+0.193) \\
         case8 & 0.426 (+0.423) & 0.446 (-0.031) & 0.425 (+0.263) & 0.491 (+0.741) \\
         case9 & 1.347 (+0.084) & 1.588 (+0.073) & 1.410 (-0.018) & 1.379 (+9341.530) \\
         case10 & 0.193 (+0.151) & 0.097 (+1.978) & 0.046 (+1.758) & 0.131 (+3.360) \\
         case11 & 1.337 (-0.005) & 1.178 (+0.054) & 1.182 (+1.933) & 1.254 (+1.310) \\
         case12 & 1.772 (+0.555) & 1.734 (+0.636) & 1.630 (+1.047) & 1.720 (+1.583) \\
         \hline
         Avg. Imp. & +0.501 & +0.933 & +0.817 & +781.032 \\
    \hline 
    \end{tabular}
    }
    \\
    \vspace{0.3cm}
    \resizebox{\linewidth}{!}{
    \begin{tabular}{|c|cccc|cccc|}
        \hline
          & \multicolumn{4}{c|}{\textbf{WNS}} &\multicolumn{4}{c|}{\textbf{TNS}} \\
         & 1st + GAP-LA & 2nd + GAP-LA & 3rd + GAP-LA & \cite{Zhao25helemgr} + GAP-LA & 1st + GAP-LA & 2nd + GAP-LA & 3rd + GAP-LA & \cite{Zhao25helemgr} + GAP-LA  \\
        \hline
        case1 & -0.366 (+15.4\%) & -0.366 (+8.2\%) & -0.367 (+3.6\%) & -0.364 (+31.8\%) & -995.9 (+24.9\%) & -986.8 (+26.4\%) & -987.7 (+13.2\%) & -992.5 (+32.7\%) \\
        case2 & -0.391 (+7.06\%) & -0.387 (+8.7\%) & -0.384 (+5.6\%) & -0.389 (+11.2\%)  & -9594.3 (+7.4\%) & -9541.2 (+11.4\%) & -9478.6 (-1.4\%) & -9635.2 (+11.8\%)\\
        case3 & -57.589 (-1.2\%) & -56.475 (-1.1\%) & -57.642 (-3.1\%) & -56.905 (+63.9\%) & -520231.3 (+0.2\%) & -518380.2 (-0.7\%) & -519322.0 (-0.9\%) & -517720.5 (+40.2\%) \\
        case4 & -0.383 (-1.3\%)& -0.384 (+21.5\%) & -0.387 (+25.5\%) & -0.390 (+45.1\%) & -1678.2 (-4.2\%)& -1673.6 (+24.8\%) & -1669.5(+33.9\%) & -1735.9 (+52.8\%) \\
        case5 & -0.276 (+9.7\%) & -0.258 (+25.6\%) & -0.253 (+8.2\%) & -0.314 (+65.1\%) & -17315.2 (+15.4\%) & -17153.2 (+21.0\%) & -16941.9 (+12.4\%) & -17674.9 (+64.5\%) \\
        case6 & -0.249 (+6.4\%) & -0.246 (+21.8\%) & -0.251 (+9.0\%) & -0.248 (+68.3\%)& -52866.6 (+11.5\%) & -52649.0 (+15.7\%) & -52342.4 (+12.1\%) & -52505.1 (+37.5\%)\\
        case7 & -0.643 (+63.4\%) & -0.610 (+17.2\%) & -0.614 (+3.0\%) & -0.668 (+59.6\%) & -111.0 (+82.9\%) & -101.3 (+32.2\%) & -102.0 (+37.2\%) & -117.4 (+78.2\%) \\
        case8 & 0 (+100.0\%) & 0 (-) & 0 (+100.0\%) & 0 (+100\%) & 0 (+100.0\%) & 0 (-) & 0 (+100.0\%) & 0 (+100.0\%)  \\
        case9 & 0 (-) & 0 (-) & 0 (-) & 0 (+100\%) & 0 (-) & 0 (-) & 0 (-) &  0 (+100.0\%)\\
        case10 & -0.348 (-1.6\%) & -0.329 (+30.3\%) & -0.333 (+24.4\%) & -0.329 (+45.0\%) & -1315.0 (+3.0\%) & -1303.2 (+32.1\%) & -1289.2 (+28.6\%) & -1299.0 (+48.9\%) \\
        case11 & -0.403 (+7.0\%) & -0.352 (+23.4\%) & -0.351 (+5.8\%) & -0.366 (+73.7\%) & -27048.8 (+15.5\%) & -25870.8 (+25.8\%) & -25364.8 (+5.4\%)& -26550.1 (+54.2\%)\\
        case12 & -0.193 (+13.1\%) & -0.209 (+23.4\%) & -0.194 (+5.8\%) & -0.212 (+87.5\%) & -37883.3 (+11.0\%) & -37334.2 (+25.9\%) & -36737.6 (+7.0\%) & -37835.3 (+72.5\%)  \\
        \hline
        Avg. Imp. & +11.8\% & +18.3\%& +8.5\% & +71.3\% & +3.6\% & +5.1\% & +2.0\% & +49.4\%  \\
        \hline
        
    \end{tabular}
    }

    \vspace{0.3cm}
    \resizebox{\linewidth}{!}{
    \begin{threeparttable}
    \begin{tabular}{|c|cccc|cccc|}
        \hline
        & \multicolumn{4}{c|}{\textbf{Power}}  & \multicolumn{4}{c|}{\textbf{Congestion}} \\
         & 1st + GAP-LA & 2nd + GAP-LA & 3rd + GAP-LA & \cite{Zhao25helemgr} + GAP-LA & 1st + GAP-LA & 2nd + GAP-LA& 3rd + GAP-LA & \cite{Zhao25helemgr} + GAP-LA   \\
        \hline
        case1 & 0.646078 (+0.0\%) & 0.646074 (+0.0\%)  & 0.646078 (-0.0\%) & 0.646082 (+0.0\%) & 6655094 (-13.8\%) & 6615019 (-12.1\%) & 6684393 (+16.5\%) & 6668336 (-14.7\%) \\
        case2 & 3.05347 (+0.0\%) & 3.05335 (-0.0\%) & 3.05335 (+0.0\%) & 3.05345 (-0.0\%) & 22926581 (-7.5\%) & 22565487 (-10.2\%) & 22404691 (+11.8\%) & 24564053 (-7.9\%) \\
        case3 & 2.93845 (+0.1\%) & 2.93804 (+0.1\%) & 2.93750 (+0.0\%) & 2.93792 (+1.4\%) & 13512436 (-1.5\%) & 14294566 (-4.4\%) & 13539305 (+13.7\%) & 13774919 (-6.2\%) \\
        case4 & 0.143804 (-0.5\%) & 0.143590 (+1.1\%) & 0.143594 (+1.1\%) & 0.143795 (+1.2\%) & 2691325 (+10.9\%) & 2667440 (-7.6\%) & 2691663 (+22.4\%) & 2695234 (-36.7\%) \\
        case5 & 7.58974 (+1.3\%) & 7.57485 (+2.1\%) & 7.55822 (-0.1\%) & 7.57887 (+3.2\%) & 67951941 (-23.5\%) & 68592745 (-39.8\%) & 68488719 (+29.8\%) & 68206894 (-54.6\%) \\
        case6 & 23.1439 (+1.3\%) & 23.1022 (+2.0\%) & 23.0783 (-0.1\%) & 23.0946 (+2.6\%) & 246164776 (-3.9\%) & 245829220 (-28.0\%) & 239426580 (+17.4\%) & 252417493 (-30.7\%) \\
        case7 & 0.156168 (-0.0\%) & 0.156157 (-0.0\%) & 0.156148 (-0.0\%) & 0.156158 (-0.0\%) & 4382560 (-0.8\%) & 4410642 (+5.2\%)& 4383521 (+0.1\%) & 4417159 (-2.5\%) \\
        case8 & 0.304879 (-0.1\%) & 0.304863 (-0.0\%) & 0.304885 (-0.0\%) & 0.304852 (-0.0\%) & 21601536 (+4.3\%) & 22622941 (-6.2\%) & 21556732 (-4.8\%) & 24942228 (+5.5\%) \\
        case9 & 0.136300 (-0.0\%) & 0.136306 (-0.0\%) & 0.136301 (-0.0\%) & 0.136310 (+0.2\%) & 13167451 (+6.4\%) & 15576558 (+4.7\%) & 13799305 (-1.2\%) & 13475394 (-3.1\%) \\
        case10 & 0.143850 (-0.6\%) & 0.143695 (+0.8\%) & 0.143565 (+1.1\%) & 0.143783 (+1.2\%) & 2074929 (+7.0\%) & 2077647 (+2.1\%) & 2059059 (+5.1\%) & 2113216 (-24.6\%) \\
        case11 & 8.19537 (+0.9\%) & 8.14163 (+1.7\%) & 8.15759 (+0.1\%) & 8.17356 (+5.3\%) & 68079290 (-29.0\%) & 69214810 (-39.2\%) & 68307458 (+48.3\%) & 68304067 (-60.2\%) \\
        case12 & 24.0157 (+1.1\%) & 23.9685 (+1.8\%) & 23.9422 (-0.1\%) & 23.9641 (+2.2\%) & 235753772 (+2.9\%) & 241013713 (-12.6\%) & 236685305 (+29.4\%) & 240225253 (-22.3\%) \\
        \hline
        Avg. Imp. & +1.1\% & +1.8\% & -0.1\% & +2.6\% & -4.6\% & -20.9\% & +26.1\% & -27.7\% \\
        \hline
        
    \end{tabular}
    \begin{tablenotes}
        \item $\dag$ The numbers in the brackets denote the improvement w.r.t the original results of 1st, 2nd, 3rd winners and~\cite{Zhao25helemgr}, respectively, by replacing their layer assignment with GAP-LA. 
        \item \quad Positive numbers indicate improvement in metrics and negative numbers indicate a degradation.
    \end{tablenotes}
    \end{threeparttable}
    }

\end{minipage}
\end{figure*}

We implemented our GPU-accelerated layer assignment algorithm using \texttt{C++} and \texttt{CUDA} on top of \texttt{HeLEM-GR}~\cite{Zhao25helemgr} and \texttt{HeteroSTA}~\cite{gputimertcad23, heteroexcepticcad24, HeteroSTAWebsite, guo2025heterosta}, and evaluated the results on benchmarks from the ISPD25 contest~\cite{ISPD2025} released by NVIDIA. 
The ISPD25 contest released 12 industrial designs synthesized with the modified Nangate 45nm open library technology node and specified an official evaluation flow to assess routing solution performance, as detailed in Section~\ref{sec:prelim}.
The details of 12 designs and the Nangate 45nm library are shown in Table~\ref{table:details} and~\ref{table:Nangate45}, respectively.
Note that the modified Nangate 45nm library holds smaller resistance and capacitance on high layers.
All experiments were conducted on a Linux server equipped with a 32-core Intel Xeon Platinum 8358 CPU@2.60GHz and an NVIDIA A800 GPU. 

To match the contest evaluation setup, we used the official evaluator for overflow assessment and \texttt{OpenROAD} for evaluating TNS, WNS, and power consumption utilizing 8 CPU threads.
\subsection{Main Comparison with Top-3 Winners in the Contest}
We obtained the top-3 contest winners' results and reproduced their experiments on our hardware for comparison.
Table~\ref{table:QoR and RT} presents a comparison of quality score, runtime and via count against the ISPD 2025 contest's top-3 winners. 
Our framework achieved the highest quality score in 9 out of 12 benchmark cases, with comparable performance in the remaining 3 cases.
While our static timing analysis engine incurs a 25\% runtime overhead due to parsing netlist (.v) and Synopsys design constraints (.sdc) files, the optimization method remains efficient and effective, demonstrating competitive performance relative to the top-3 winners.
Our GPU-accelerated layer assignment method shows over 2.18$\times$ speedup on average, and over 2.6$\times$ speedup on large cases over the multi-threaded version on CPU.
As Table~\ref{table:QoR and RT} shows that high layers like metal9 and metal10 have significant lower resistance and capacitance, more via usage implies more radical optimization towards timing. 
The 2nd winner tries to use the fewest resources on high layers and the 3rd winner tries to use the most resources, while our framework follows a more balanced way.

Table~\ref{table:metrics} presents the WNS, TNS, power consumption, and congestion results of four routers across 12 benchmark cases.
The 2nd-place winner produces solutions with favorable performance and low congestion, while the 3rd-place winner demonstrates better performance at the expense of higher congestion.
The 1st-place winner achieves a more balanced trade-off among metrics~\cite{ISPD2025}.
Our method delivers superior performance with acceptable congestion increase, which can be attributed to the critical net analysis-based net ordering and batching strategy.
As shown in Table~\ref{table:metrics}, compared to the 1st-place winner, our router achieves 9.9\% better WNS, 4.5\% better TNS, 1.1\% lower power consumption, and 3.8\% higher congestion. 
Versus the 2nd-place winner, we achieve 0.3\% better WNS, 5.4\% better TNS, 1.5\% lower power consumption, and 15.5\% higher congestion. 
In contrast, the 3rd-place winner shows only 0.4\% lower power consumption but suffers 4.4\% worse WNS, 2.0\% worse TNS, and 35.2\% higher congestion compared to our approach.

\subsection{Comparison with Top-3 Winners and Congestion-driven Router Using Our Method\label{subsec:using}}
To better illustrate the effectiveness of our proposed method, we compress the 3D routing solution given by the top-3 winners into 2D net topology and apply our method to them.
Table~\ref{table:replace} shows the comparison of overall quality score, WNS, TNS, power and congestion with their original flow.
The runtime comparison is omitted here since we only have the binaries and cannot extract their layer assignment runtime.
As the contest targets at optimizing the overall quality scores, using GAP-LA on top of the 2D net topology from the top-3 winners can lead to 0.501, 0.933, and 0.817 better quality scores, respectively. 
In more detail, our method gains 11.8\%, 18.3\%, 8.5\% better WNS, 3.6\%, 5.1\%, 2.0\% better TNS than their original flow, respectively.
Although we sacrifice 4.6\% and 20.9\% congestion compared with the original flow of 1st and 2nd winner, we improve the performance a lot, given the affect that the quality scores from the contest emphasize more on timing metrics.
Compared with the 3rd winner, our method can achieve better performance with less congestion.
As for the congestion-driven layer assignment method~\cite{Zhao25helemgr}, our method achieves 71.3\% better WNS, 49.4\% better TNS and 2.6\% less power with 27.7\% more congestion on average, which emphasizes the importance of leveraging performance-aware strategy on our benchmarks.

\subsection{Comparison with timing-aware method~\cite{Liu22TALA}}
There are several works focusing on timing-aware layer assignment~\cite{Zhang20MiniDelay, Jiang22LASVR, Liu22TALA}, among which~\cite{Liu22TALA} is the most recent one.
To detail our advantages, we compare WNS, TNS, power and congestion with~\cite{Liu22TALA} on three representative cases in Table~\ref{table:compare with TALA}.
We sincerely appreciate the help from the authors of~\cite{Liu22TALA} to run the evaluation and obtain the permission to show these results.
However, the comparison is definitely unfair due to design imcompatibility and different optimization objectives.
The objective of~\cite{Liu22TALA} is to minimize the timing cost under the congestion constraint instead of optimizing the slack with acceptable overflow in our problem setting.
Therefore, the comparison is just for reference and does not downplay the effectiveness of~\cite{Liu22TALA}.
\begin{table*}
\centering
\begin{threeparttable}
    \caption{Comparison of WNS, TNS, Power and Congestion with timing-aware layer assignment method~\cite{Liu22TALA}.\label{table:compare with TALA}}
    \begin{tabular}{|c|cc|cc|cc|cc|}
        \hline
         & \multicolumn{2}{c|}{\textbf{WNS}} & \multicolumn{2}{c|}{\textbf{TNS}} & \multicolumn{2}{c|}{\textbf{Power}} & \multicolumn{2}{c|}{\textbf{Congestion}}  \\
         & ~\cite{Liu22TALA} & Ours & ~\cite{Liu22TALA} & Ours & ~\cite{Liu22TALA} & Ours & ~\cite{Liu22TALA} & Ours \\
        \hline
        case1 & -2.613 & -0.365 & -7789.6 & -997.9 & 0.646533 & 0.646076 & 1144690923663 & 6662339 \\ 
        case4 & -0.915 & -0.387 & -4672.2 & -1711.6 & 0.149762 & 0.14383 & 59494669 & 67934233 \\ 
        case7 & -4.421 & -0.645 & -5066.9 & -111.5 & 0.156124 & 0.15614 & 7099591 & 4371048 \\ 
        \hline
    \end{tabular}
    \begin{tablenotes}
        \footnotesize
        \item We sincerely appreciate the help from the authors of~\cite{Liu22TALA} to run the evaluation and obtain the permission to show these results. 
              The comparison is just for reference and does not downplay the effectiveness of~\cite{Liu22TALA} due to design incompatibility and different optimization objectives.
    \end{tablenotes}
\end{threeparttable}
\end{table*}

\subsection{Ablation Studies}
\begin{table*}[t]
    \centering
    \caption{Ablation study on our critical net analysis-based net ordering and batching strategy and look ahead strategy.}
    \label{table:w/o net ordering}
    \resizebox{\linewidth}{!}{
        \begin{tabular}{|cc|cccccccccccc|c|}
    \hline
        & & case1 & case2 & case3 & case4 & case5 & case6 & case7 & case8 & case9 & case10 & case11 & case12 & Avg. \\
        \hline
        \multirow{3}{*}{\textbf{WNS}} & Ours & \textbf{-0.365} & \textbf{-0.388} & \textbf{-56.25} & \textbf{-0.387} & \textbf{-0.327} & \textbf{-0.247} & \textbf{-0.645} & \textbf{0.000} & \textbf{0.000} & \textbf{-0.353} & \textbf{-0.404} & -0.199 & \textbf{-4.961} \\
        & w/o ordering & -0.385 & -0.404 & -60.415 & -0.488 & -0.293 & -0.272 & -1.458 & \textbf{0.000} & -10.501 & -0.413 & -0.430 & -0.214 & -6.273 \\
        & w/o ahead & -0.368 & -0.398 & -56.316 & -0.426 & -0.330 & -0.251 & -0.791 & \textbf{0.000} & \textbf{0.000} & -0.361 & -0.405 & \textbf{-0.198} & -4.987 \\
        \hline
        \multirow{3}{*}{\textbf{TNS}} & Ours & \textbf{-997.9} & \textbf{9602.9} & \textbf{-515014.3} & \textbf{-1711.6} & \textbf{-17619.7} & \textbf{-52800.0} & \textbf{-111.5} & \textbf{0.000} & \textbf{0.000} & \textbf{-1312.2} & \textbf{-27116.0} & \textbf{-38106.6} & \textbf{-55366.1} \\
        & w/o ordering& -1065.1 & -9768.3 & -549035.6 & -2167.1 & -18133.0 & -56966.8& -387.3 & \textbf{0.000} & -10903.8 & -1626.9 & -28959.3 & -39449.7 & -59874.4 \\
        & w/o ahead & -1061.4 & -10036.4 & -515775.7 & -1839.4 & -18161.2 & -53875.2 & -194.1 & \textbf{0.000} & \textbf{0.000} & -1382.5 & -27565.8 & -38441.1 & -55694.4 \\
        \hline
    \end{tabular}
    }
\end{table*}
To evaluate the effectiveness of our proposed strategies, we compare the WNS and TNS result of our solution on each design with  those that do not use these strategies.
\begin{table}
    \caption{Number of critical paths and critical nets on several selected cases.\label{table:critical}}
    \centering
    \begin{tabular}{|c|c|c|c|c|}
        \hline
        Case & \# Crit. Paths & \# Crit. Nets &  $N_{net}$ & Crit. Ratio \\
        \hline
        1 & 1585 & 360 & 123900 & 0.291\% \\
        \hline
        2 & 1098 & 458 & 736883 & 0.062\% \\
        \hline
        3 & 4102 & 810 & 136120 & 0.595\% \\
        \hline
        7 & 312 & 91 & 105924 & 0.086\% \\
        \hline
        8 & 630 & 9 & 768239 &  0.001\% \\
        \hline
        9 & 4034 & 816 & 135814 & 0.601\% \\
        \hline
    \end{tabular}
\end{table}

\begin{figure}[b]
    \centering
    \subfloat[]{
        \includegraphics[width=0.45\linewidth]{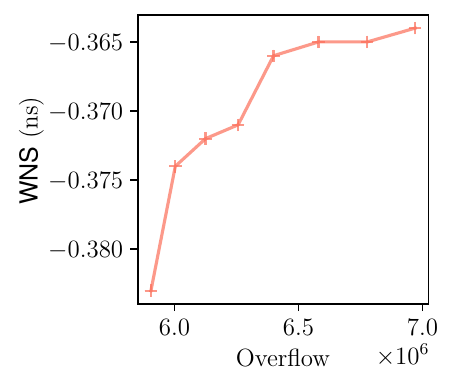}
    }
    \subfloat[]{
        \includegraphics[width=0.45\linewidth]{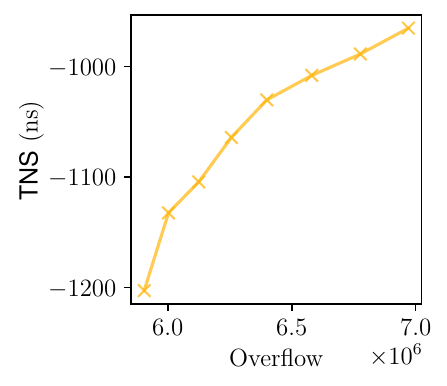}
    }
    \caption{Overflow, (a) WNS and (b) TNS in case 1 under different delay and capacitance weights.\label{fig:weights}}
\end{figure}

Table~\ref{table:w/o net ordering} shows that our framework achieves better WNS results in 11 of the 12 ISPD2025 benchmark cases and better TNS results in all 12 cases. 
The net ordering and batching strategy enables more precise optimization of critical nets, significantly outperforming the naive strategy that ignores timing analysis information. 
This advantage is particularly evident in cases sensitive to critical net delay and capacitance, such as case 9.
The look-ahead strategy also proves crucial in certain cases, including case 4.

Table~\ref{table:critical} presents the number of critical paths and nets for several selected cases.
The results show that only a small fraction of nets (less than 1\% of all nets) in the netlist play a significant role in timing closure.
 In case 8, focusing optimization efforts on the 9 critical nets with high timing cost weights is sufficient to achieve zero slack, demonstrating the effectiveness of our critical net selection strategy.

Figure~\ref{fig:weights} illustrates the variations in overflow, WNS, and TNS as the delay and capacitance weights increase.
Our framework generates solutions with improved timing performance but higher congestion, indicating that the method can adapt to different optimization priorities to meet diverse design requirements.

\section{conclusion}
\label{sec:conclu}

In this paper, we present a GPU-accelerated performance-driven layer assignment framework that generates high-quality routing solutions from 2D routing results.
The framework integrates a static timing analysis engine to schedule optimal net ordering and task batching for effective performance optimization. 
To address the delay cost optimization of large-scale designs with up to 10 million nets, we design GPU kernels for layer assignment using a look-ahead strategy. 
Compared with the top-3 winners of the contest, our router achieves 0.3\%–9.9\% better WNS and 2.0\%–5.4\% better TNS on average with acceptable runtime overhead.
    We apply our layer assignment framework to the 2D net topology of the top-3 winners and congestion-driven router~\cite{Zhao25helemgr}.
    Compared with the original flow of top-3 winners, our method gains 0.501-0.933 better quality score, 8.5\%-18.3\% better WNS, 2.0\%-5.1\% better TNS on average, respectively.
\section{Acknowledgments}
Sincere thanks are given to Prof. Genggeng Liu and his student Zepeng Li for their help to run the evaluation of~\cite{Liu22TALA} and permission to show these results.

\bibliographystyle{IEEEtran}
\bibliography{ref.bib}
\begin{IEEEbiography}
[{\includegraphics[width=1in,height=1.25in,clip,keepaspectratio]{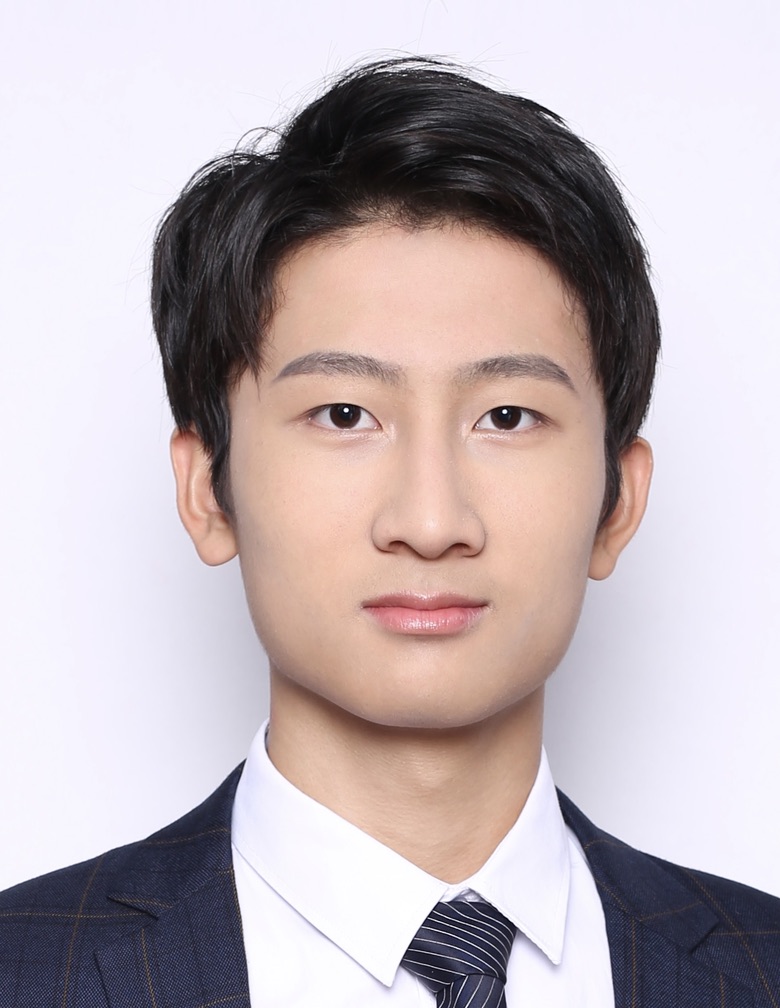}}]{Chunyuan Zhao}
    is a second-year PhD student in the School of Integrated Circuits at Peking University, advised by Prof. Yibo Lin.
    Before that, he obtained his B.S. degree in the College of Engineering at Peking University, in 2024. 
    His research interests include data structure, algorithm and GPU-acceleration in electronic design automation (EDA). 
    He won the first place in the ISPD contest held by NVIDIA in 2024 and 2025.
    He is also a recipent of the Best Paper Award Nomination in ICCAD 2025.
\end{IEEEbiography}
\vspace{-0.2cm}

\begin{IEEEbiography}
[{\includegraphics[width=1in,height=1.25in,clip,keepaspectratio]{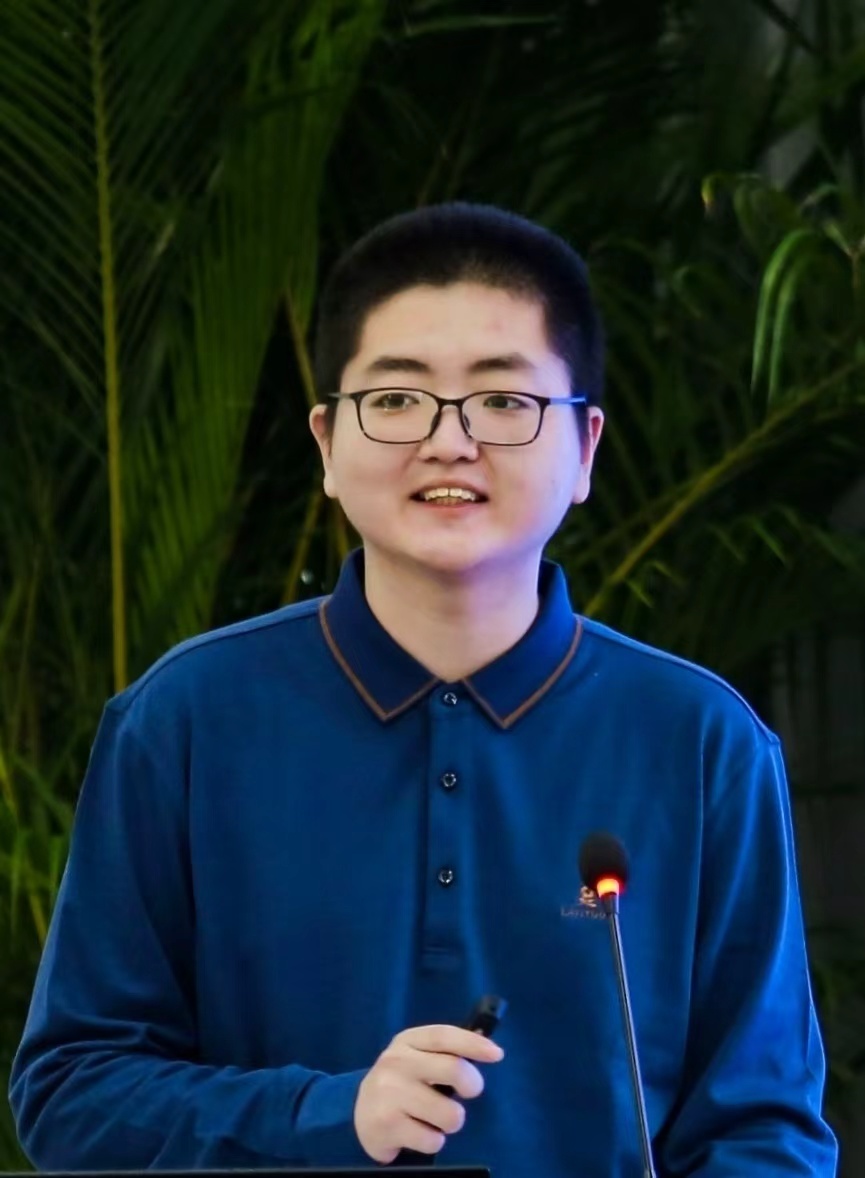}}]{Zizheng Guo}
is a 4th year Ph.D. candidate at Peking University advised by Prof. Yibo Lin. 
His research interests include data structures, algorithm design and GPU acceleration for combinatorial optimization problems in physical design automation. 
He received his B.Sc. degree in computer science from Peking University in 2022.
 He won several first places in ICCAD CAD contest 2024, ICCAD CADathlon contest 2024, EDAthon 2022, and ACM Student Research Competition Grand Finals 2022. 
 His works are nominated best paper several times at DAC 2025, ICCAD 2024, ASP-DAC 2024, and ISEDA 2024.
\end{IEEEbiography}
\vspace{-0.2cm}

\begin{IEEEbiography}
[{\includegraphics[width=1in,height=1.25in,clip,keepaspectratio]{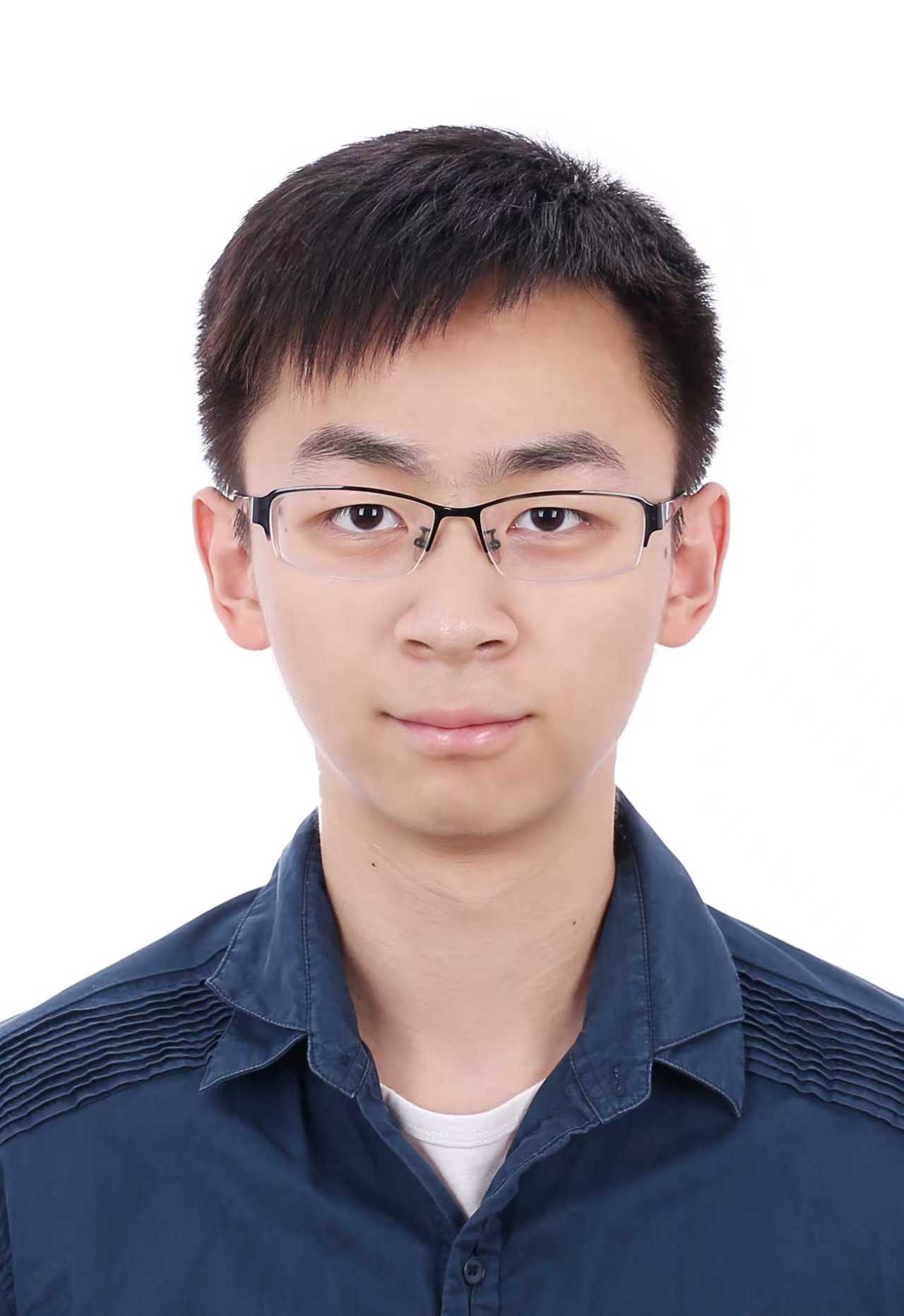}}]{Zuodong Zhang}
received his B.S. degree in Electronic Information Engineering from Beihang University in 2018, and his Ph.D. degree from the School of Integrated Circuits at Peking University in 2023. 
He is currently working on digital backend EDA algorithms.
\end{IEEEbiography}
\vspace{-0.2cm}

\begin{IEEEbiography}[{\includegraphics[width=1in,height=1.25in,clip,keepaspectratio]{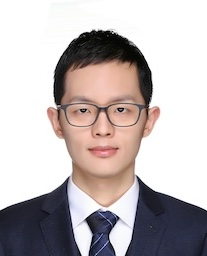}}]{Yibo Lin}
received the B.S. degree in Microelectronics from Shanghai Jiaotong University in 2013. He obtained his Ph.D. degree in Electrical and Computer Engineering from the University of Texas at Austin in 2018. He currently is an Associate professor in the School of Integrated Circuits at Peking University. His research interests include physical design, machine learning applications, and heterogeneous computing in VLSI CAD. He is a recipient of the Best Paper Awards at preimier EDA/CAD journals/conferences like TCAD, DAC, DATE, ISPD, etc.
\end{IEEEbiography}

\vfill

\end{document}